# HI and CO Observations of the Circumnebular Envelopes of Planetary Nebulae

G. T. Gussie

Department of Physics

University of Tasmania

GPO Box 252C

Hobart, Tasmania, Australia 7001

internet: Grant.Gussie@phys.utas.edu.au

A. R. Taylor

Department of Physics and Astronomy

University of Calgary

2500 University Drive N.W.

Calgary, Alberta, Canada T2N 1N4

internet  russ@ras.ucalgary.ca

ABSTRACT

$\lambda$ = 21 cm spectral-line and continuum observations of ten compact planetary nebulae have been made with the Very Large Array. Circumnebular atomic hydrogen is identified in the planetary nebulae SwSt 1, M3-35, and possibly in Hu 2-1. The possibility of an inverse correlation between the optical depth of circumnebular atomic hydrogen and the linear radius of the ionised nebula is investigated, but not shown to be statistically significant. Observations of $J = 2 \rightarrow 1$ CO emission for 22 planetary nebulae and $J = 3 \rightarrow 2$ CO emission for 16 planetary nebulae taken with the James Clerk Maxwell Telescope are presented. $J = 2 \rightarrow 1$ CO emission is detected in the planetary nebulae M1-59, and previous CO detections in IC 5117, M1-16, and NGC 6302 are confirmed. $J = 3 \rightarrow 2$ CO emission is detected in the planetary nebulae IC 5117, and NGC 6302.

1. INTRODUCTION

A complete understanding of the structure, kinematics, composition, and evolution of planetary nebulae can only be obtained if knowledge of nebulae's ionised, molecular, and atomic gas as well as the solid material is obtained. Study of the ionised gas of planetary nebulae began with the advent of modern astronomy, and consequently has a long history, both at optical and radio wavelengths. More recently, infrared observations of nebular and circumnebular dust grains have provided information of the nebulae's dust component. Spectral line observations of circumnebular molecules have also been made in recent years, and have proven fruitful, especially observations of CO (e.g. Huggins & Healy 1989). Observations of the neutral atomic component of planetary nebulae (hereafter PN) have also been made. Both atomic oxygen and atomic sodium have been observed in planetary nebulae. Atomic carbon, oxygen, and sodium have all been observed in PNe (Bachiller et al. 1994; Dinerstein 1991; Dinerstein, Haas, & Werner 1991). Surprisingly difficult to observe is however the most common of astronomical atomic species, HI. Although HI can be detected via its $\lambda = 21$ cm line, observations of HI associated with PN have met with only limited success, despite a number of careful investigations (e.g., Thompson & Colvin 1970; Zuckerman, Terzian, & Silverglate 1980; Rodríguez et al. 1985; Altschuler et al. 1986; Schneider et al. 1987). Observations of the $\lambda = 21$ cm line of HI have however been successful in detecting circumnebular atomic material in a few cases when the challenges of sensitivity and interstellar confusion were overcome (e.g., Taylor, Gussie, & Pottasch 1990, hereafter TGP).

The TGP survey revealed a possible inverse correlation between the linear radius of the ionised nebula and the optical depth of circumnebular HI, suggesting that the growth of the ionised nebula results in the consumption of surrounding atoms. This is consistent with the Huggins & Healy (1989) result that the ionised mass of PNe grows at the expense of the molecular (CO) envelope. The HI data of TGP was however insufficient for the relationship between atomic mass and ionised radius to be demonstrated, as the data were few and the uncertainties in distance were large.

In order to determine whether the HI column density of the PN circumnebular envelopes decreases with increasing ionised radius in compact PN (as suggested by the TGP survey results) an additional sample of compact planetary nebulae were selected for observation. Since the ionised linear radii of the planetary nebulae are often difficult to estimate – due to the difficulty in determining PN distances – compact nebulae were distinguished from larger but more distant objects by the higher radio continuum brightness temperature of the former. The radio continuum brightness temperature of a large sample of planetary nebulae was calculated from the angular sizes and $v = 5$ GHz radio continuum flux densities as listed in a number of sources (notably Aaquist & Kwok 1990). Those nebulae with $T_{B(5GHz)} \geq 500$ K were selected for further consideration. Of these nebulae, only those with known systemic velocities (Schneider et al. 1983) that differed significantly from the velocity of Galactic hydrogen emission (Weaver & Williams 1974) were further considered, as was required to avoid confusion with interstellar hydrogen. Previously published evidence of a neutral envelope (e.g. from $H_2$ or CO observations) and a high Galactic latitude were also considered to be desirable in sample nebulae, but were not actively selected for. Ten nebulae were chosen that fit the above criteria and were suitably placed in the sky for observation with the Very Large Array (VLA). These nebulae formed the sample for the present HI absorption survey. In addition, a concurrent detection survey of $J = 2 \rightarrow 1$ and $J = 3 \rightarrow 2$ CO emission in planetary nebulae was conducted with the James Clerk Maxwell Telescope (JCMT). These observations were to increase the number of known CO sources among planetary nebulae with a more sensitive survey than previous attempts. Source selection was via similar criteria as the HI survey, although high radio continuum brightness was not a high priority. An attempt was made to create as much overlap as possible in the two survey samples, but the limitations of time and the different locations of the two telescopes (New Mexico vs. Hawaii) made duplication of the two samples impractical.

## 2. THE HI OBSERVATIONS

Data for the HI line survey were collected with the VLA on four separate days: on 5 March 1989 and 30 April 1989 while the telescope was in B configuration; and on 28 July 1989 and 31 July 1989 while the telescope was in C configuration.

The observations were made with 63 spectral-line channels with on-line Hanning smoothing. The data averaging time was 60 seconds. The effective spectral resolution (after Hanning smoothing) was $\Delta \nu = 12.207\,\mathrm{kHz}$, corresponding to a Döppler velocity resolution of $\Delta v = 2.56\,\mathrm{km\,s^{-1}}$. Twenty-six of the twenty-seven VLA antennae were used on each day of the observations; there being one antenna inoperative on each day of observations. The spectrometer configuration provided 127 spectral-line channels but only the central 50% of available channels were used since the on-line computers of the VLA were at that time incapable of accepting the large data throughput of 127 channels from the 26 antennae.

The expansion velocity of the circumnebular HI in each detected nebula is determined from the formula $v_{exp(HI)} = v_{max} - v_{HWHM} - v_{LSR}$, where $v_{max}$ is the position of the absorption line's maximum depth, $v_{HWHM}$ is the half width at half maximum of a Gaussian fit to the observed absorption line, and $v_{LSR}$ is the velocity of the ionised nebula with respect to the local standard of rest. The $v_{HWHM}$ is subtracted from $v_{max}$ to correct for the finite velocity resolution of the observations in a first order manner.

## 3. RESULTS OF HI OBSERVATIONS

Circumnebular atomic hydrogen is detected in absorption in SwSt 1 (figure 1a). To increase the signal to noise ratio of these spectra adjacent channels of the original 63-channel spectra were averaged to create 31-channel spectra. The spectral resolution of the resultant spectra is $\Delta v = 5.12\,\mathrm{km\,s^{-1}}$, which still provides about 4 velocity resolution elements across the observed absorption line. SwSt 1 is believed to be very young, with a near circular morphology and an ionised linear radius of < 1" (Maciel 1984), giving a calculated linear radius of only 0.0038 pc. The integrated optical depth of circumnebular HI in SwSt 1 is $\int \tau \, dv$ = $(8.0 \pm 0.6) \, \mathrm{km \, s^{-1}}$.

The planetary nebula Hu 2-1 (figure 1b) is found to possess possible circumnebular HI with an integrated optical depth of $\int \tau dv$ = $(0.68 \pm 0.09) \, \mathrm{km \, s^{-1}}$. However, because of the unusual narrowness of the HI absorption line, it is possible that the absorption is an unusually large noise peak which unfortunately lies at the velocity expected for circumnebular HI. The expansion velocity of the circumnebular HI is found to be $v_{exp}$ = $(5 \pm 2) \, \mathrm{km \, s^{-1}}$. The relatively low expansion velocity casts further doubt as to the validity of the detection. However, the detection is significant at greater than the $5\sigma$ level and occurs at precisely the expected velocity. This detection requires reconfirmation. Hu 2-1 is also a very small nebula (Maciel 1984), but is already displaying a deviation from spherical symmetry perhaps indicative of a future bilobate morphology (Aaquist & Kwok 1990).

Circumnebular HI is observed in the planetary nebula M 3-35 (figure 1c) with an integrated optical depth of $\int \tau dv$ = $(2.5 \pm 0.3) \, \mathrm{km \, s^{-1}}$. The expansion velocity of the HI is found to be $(22.0 \pm 3.5) \, \mathrm{km \, s^{-1}}$. M 3-35 is known to be another small and circular PN (Aaquist & Kwok), and likely to be quite young.

The nebula M 1-78 was included in this sample because of its previous inclusion in the Perek-Kohoutek PN catalogues, and because it fulfils (quite admirably) the source selection criteria given above. However, subsequent investigation into the classification of this nebula has shown (Gussie 1994) that M 1-78 is in all probability *not* a PN, but rather is likely to be an ultra-compact HII region with an evolved central star. Atomic hydrogen was nevertheless found near the systemic velocity of the nebula M 1-78 (figure 1d). The bulk of the HI is found at velocities that are red-shifted with respect to the systemic velocity of the ionised nebula, not blue-shifted as is the HI in the nebulae known to exhibit circumnebular HI absorption. The observed HI gas is therefore quite probably interstellar in origin. Nevertheless, the possibility exists that this gas is somehow associated with the nebula since it occurs at the same velocity as the nebula's observed CO emission (Zuckerman et al. 1977). The integrated optical depth of HI is found to be $\int \tau dv = (0.07 \pm 0.02) \,\mathrm{km\,s^{-1}}$.

The results of the HI observations are summarised in table 1.

## 4. THE CO OBSERVATIONS

Both the $J = 2 \rightarrow 1$ CO line at $v = 230\,\mathrm{GHz}$ and the $J = 3 \rightarrow 2$ CO line at $v = 345\,\mathrm{GHz}$ were observed with the JCMT. Observing dates were 29-30 May 1989, 7-8 May 1990, and 4-7 May 1991. Observations at $v = 230\,\mathrm{GHz}$ were conducted with the JCMT's A1 receiver while the $v = 345\,\mathrm{GHz}$ observations were conducted with the B1 receiver. Attached to both receivers was the "Acousto-optical Spectrometer, Canadian" (AOSC), which gave a spectral velocity resolution of $\Delta v = 0.43\,\mathrm{km\,s^{-1}}$ at $v = 230\,\mathrm{GHz}$ and $\Delta v = 0.29\,\mathrm{km\,s^{-1}}$ at $v = 345\,\mathrm{GHz}$. The JCMT's angular resolution is $\Delta\theta = 22"$ at $v = 230\,\mathrm{GHz}$ and $\Delta\theta = 15"$ at $v = 345\,\mathrm{GHz}$.

Sky subtraction was conducted by position switching with a throw distance of 10' in azimuth. The telescope's tracking is reported to be accurate to within $\pm 1"$ over a period of one hour. The pointing accuracy of the JCMT is considered to be good to within $\Delta\theta_{RMS} = 2"$. The pointing accuracy of the telescope was tweaked every hour or so by performing a five-point correction. The telescope is believed to have been pointed to an accuracy of better than 4" during the observations because pointing corrections determined by the five-point observations were all < 3". Therefore pointing errors are not believed to have significantly affected the flux measurements.

The accuracy of the calibration of the A1 receiver is generally considered to be accurate to within 15%. The B1 receiver's calibration is considered accurate %%to within 20%. The errors in the antenna temperature measurements only reflect the internal noise of the spectra and do not include any errors in calibration.

In accordance with the tradition of single-dish radio spectrometry, the resulting spectra are presented in units of main beam brightness temperature, measured in K. Conversion of antenna temperature $T_{MB}$ to flux in units of Jansky is performed by multiplication of $T_{MB}$ by the conversion factor $15.6\,\mathrm{Jy\,K^{-1}}$.

The expansion velocity of the carbon monoxide envelope is measured from the CO line profile as the difference between the LSR velocity of the red-shifted edge of the CO emission line and the LSR velocity of the ionised nebula. This method assumes that the red-shifted edge of the emission line defines the maximum radial expansion velocity of the gas, but does not assume that the blue-shifted emission is unaffected by self-absorption or by the absorption of the ionised nebula.

## 5. RESULTS OF CO OBSERVATIONS

$J = 2 \rightarrow 1$ CO emission was detected in IC 5117 by Huggins & Healy (1989). We also observe $J = 2 \rightarrow 1$ CO emission, with a peak antenna temperature of $T_{MB(peak)} = (0.7 \pm 0.1)\,\mathrm{K}$. The antenna temperature integrated over the velocity width of the line is $\int T_{MB}\,dv = (6 \pm 1)\,\mathrm{K\,km\,s^{-1}}$. The $J = 2 \rightarrow 1$ CO emission of IC 5117 is centered at a velocity of $v_{LSR(CO)} = (\text{-}8 \pm 1)\,\mathrm{km\,s^{-1}}$. This velocity is slightly red-shifted with respect to the systemic velocity of the ionised gas, which is $v_{LSR(ion)} = (\text{-}12.0 \pm 1.3)\,\mathrm{km\,s^{-1}}$ (Schneider et al. 1983). The expansion velocity is found to be $v_{exp(CO)} = (17 \pm 2)\,\mathrm{km\,s^{-1}}$. This expansion velocity is within the range of the expansion velocities measured for the ionised gas of this nebula (Weinberger 1989). $J = 3 \rightarrow 2$ CO emission is also detected with a peak antenna temperature $T_{MB(peak)} = (1.5 \pm 0.3)\,\mathrm{K}$. The integrated antenna temperature is $\int T_{MB}\,dv = (8.7 \pm 0.2)\,\mathrm{K\,km\,s^{-1}}$. The emission is centered at $v_{LSR(CO)} = (\text{-}16 \pm 2)\,\mathrm{km\,s^{-1}}$, or about $\text{-}4\,\mathrm{km\,s^{-1}}$ away from the systemic velocity of the ionised gas as listed in Schneider (1983) and $8\,\mathrm{km\,s^{-1}}$ from the center of the $J = 2 \rightarrow 1$ emission. The expansion velocity indicated by the $J = 3 \rightarrow 2$ CO line is $v_{exp(CO)} = 12 \pm 2\,\mathrm{km\,s^{-1}}$, consistent with the ionised expansion velocities (Weinberger 1989). There consequently seems to be some kinematic differences between the $J = 3 \rightarrow 2$ and $J = 2 \rightarrow 1$ CO emission lines. This, along with the slight red-shift of the $J = 2 \rightarrow 1$ CO emission line, may indicate an asymmetric velocity/temperature/density structure within the nebular envelope or that the $J = 2 \rightarrow 1$ line is undergoing significant self-absorption. Asymmetry is however not indicated by continuum maps of the nebula (Aaquist & Kwok) which reveal an smoothly elliptical ionised PN.

$J = 2 \rightarrow 1$ CO emission is detected in M 1-59 with a peak antenna temperature of $T_{MB(peak)} = (0.038 \pm 0.009)$ K. The integrated antenna temperature is $\int T_{MB}\, dv = (0.50 \pm 0.08)$ K km s$^{-1}$. The CO emission is centered at a velocity of $v_{LSR(CO)} = (114 \pm 1)$ km s$^{-1}$, similar to the systemic velocity of the ionised gas of $v_{LSR(ion)} = (114.2 \pm 11.4)$ km s$^{-1}$ (Schneider et al. 1983). The expansion velocity of the CO in M 1-59 is found to be $v_{exp(CO)} = (13 \pm 2)$ km s$^{-1}$, also similar to the ionised gas expansion velocity of $v_{exp(ion)} = (13 \pm 1)$ km s$^{-1}$ listed by Weinberger (1989). M 1-59 is a larger PN than those heretofore examined, with a radius of 2.5″ and a distinctly bilobate appearance (Aaquist & Kwok 1990).

Peak $J = 2 \rightarrow 1$ CO antenna temperature in the direction of M 1-16 is $T_{MB(peak)} = (0.97 \pm 0.08)$ K. Integrated antenna temperature is $\int T_{MB}\, dv = (23.0 \pm 0.8)$ K km s$^{-1}$. The $J = 2 \rightarrow 1$ CO emission line is centered at a velocity of $v_{LSR(CO)} = (42 \pm 1)$ km s$^{-1}$, which is consistent with the systemic velocity of the ionised gas, $v_{LSR(ion)} = (31.9 \pm 25.0)$ km s$^{-1}$ listed by Schneider et al. (1983). Because of the large uncertainty in the ionised gas systemic velocity, the expansion velocity of the molecular gas of this nebula could not be reliably determined using its value. Instead, the CO expansion velocity is estimated as one half the CO emission line's Döppler width and is found to be $v_{exp(CO)} = (27 \pm 3)$ km s$^{-1}$. This value is much higher than Weinberger's (1989) ionised gas expansion velocity of $v_{exp(ions)} = 10$ km s$^{-1}$. The $J = 2 \rightarrow 1$ CO emission of M 1-16 was originally discovered by Huggins & Healy (1989). Recent CO mapping of M 1-16 (Sahai e t al. 1994) has shown multiple outflow velocities and a complex morphology of the CO envelope, and this nebula is modelled as a post-AGB binary star system.

$J = 2 \rightarrow 1$ CO emission is easily detected in NGC 6302 with a peak antenna temperature of $T_{MB(peak)} = (0.67 \pm 0.04)$ K. The integrated antenna temperature is $\int T_{MB}\, dv = (13.5 \pm 0.3)$ K km s$^{-1}$. The CO emission is centered at a velocity of $v_{LSR(CO)} = (-35 \pm 1)$ km s$^{-1}$, fairly similar to the systemic velocity of the ionised gas of $v_{LSR(ion)}$ = (-31.4±2.1) km s$^{-1}$ (Schneider et al. 1983). The expansion velocity of the CO is found to be $v_{exp(CO)}$ = (28±4) km s$^{-1}$. Weinberger does not list an ionised gas expansion velocity for this nebula, as its internal motions are too complicated for an overall expansion rate to be assigned. The line shows a double peaked profile, indicating that the CO cloud may be at least partially optically thin or that the CO is spatially resolved by the JCMT beam. Given the relatively large angular size of the ionised nebula (with an optical size of ~ 1 arc minute) makes this likely, as does a $J = 1 \rightarrow 0$ CO mass (Nyman 1989) which show emission over a similar area. This detection confirms an earlier detection of $J = 2 \rightarrow 1$ CO emission by Huggins & Healy (1989). $J = 3 \rightarrow 2$ CO emission is detected with a peak antenna temperature of $T_{MB(peak)}$ = (3.8 ± 0.9) K. The integrated antenna temperature is $\int T_{MB}\, dv$ = (8.7 ± 0.5) K km s$^{-1}$.

Strong $J = 3 \rightarrow 2$ CO emission is detected in the direction of the probable HII region M 1-78, as expected from the strength of existing $J = 1 \rightarrow 0$ CO observations (Zuckerman et al. 1977). Peak antenna temperature is a very high $T_{MB(peak)}$ = (37.2 ± 0.3) K. The integrated antenna temperature is $\int T_{MB}\, dv$ = (328 ± 1) K km s$^{-1}$. The $J = 3 \rightarrow 2$ CO emission line is centered at a velocity of $v_{LSR(CO)}$ = (-66 ± 2) km s$^{-1}$. A high-velocity wing of CO emission is also seen in this nebula at $v_{LSR}$ = (-5 ± 2) km s$^{-1}$. Both of these velocities are significantly different than the systemic velocity of the ionised gas, $v_{LSR(ion)}$ = (-76±2) km s$^{-1}$ (Gussie & Taylor 1989). However, the bulk of the CO emission is at a similar systemic velocity as the observed HI absorption (above). The relationship between the neutral gas and the ionised nebula therefore remains unclear; it is possible that they are not directly related.

The results of the CO observations are presented in tables 2 and 3. The spectra of the nebulae with detected CO emission are presented in figure 3.

6. DISCUSSION

Combining the nebulae detected in circumnebular HI absorption by TGP with the present survey and with surveys by Rodríguez & Moran (1982) and Altschuler et al. (1986) results in a total of ten planetary nebulae that currently have detected circumnebular atomic envelopes.

The optical depth of circumnebular HI (integrated over the velocity width of the absorption line) is presented as a function of the estimated ionised linear radius in figure 2. The plotted data is given in tabulated form in table 3. Statistical analysis of the data reveals that the least-squares linear regression line of

$$\log_{10}[\int \tau \, dv \,] = (\text{-0.8} \pm 0.4) \log_{10}\left[R_{ion}\right] + (\text{-1.4} \pm 0.8)$$

indicating a systematic decrease in $\int \tau \, dv$ with increasing $R_{ion}$. The correlation coefficient of the log/log data is 0.7. However, since the data are few (ten) the regression errors do not reliably indicate the true uncertainty in the regression slope. Analysis of the data with $t$-distribution statistics — appropriate for smaller sample sizes (e.g. Bhattacharyya & Johnson 1977) — indicates that there is a 95% confidence that the true %%value of the slope lies within the range -0.8 ± 1.0, indicating that the perceived inverse correlation may be spurious.

The regression analysis also does not consider the very significant uncertainty in the linear radii of the nebulae, which stems from the difficulty in determining the distance to planetary nebulae (see Pottasch 1984 for a review of distance determination methods). The uncertainty in a nebula's distance (and consequently in its linear radius) is often difficult to quantify. Only two of the planetary nebulae detected in HI have what Weinberger (1989) believes is a "reliable" distance; that is, a distance that is known to within a factor of ±50%. The other nebulae may have distance errors as large as a factor of three. A Monte Carlo investigation using simulated data has indicated that the 95% %%confidence interval of the slope is -0.8±2.2, assuming an error of a factor of three in the linear radii (arguably a worst case) and given the uncertainties in the measured optical depths. This would dictate further caution in accepting the inverse correlation between HI optical depth and linear radius. The possible inverse correlation between optical depth and ionised radius consequently remains unconfirmed, and it is not as yet clear that the ionised nebular does indeed grow at the expense of an outer atomic shell. The observed optical depths and the upper limits for the undetected nebulae do however remain consistent with this hypothesis.

The existence (or non-existence) of an inverse correlation between HI optical depth and the linear ionized radius is of interest since it provides clues as to the origin and evolution of the atomic envelopes. If the atomic gas were the remnant of a period of atomic mass loss during AGB evolution, we would expect that the atomic gas would be confined within a larger molecular cloud, since no star presently on the AGB is known to possess an atomic wind (*e.g.* Knapp & Bowers 1983). The atomic mass-loss phase must therefore be quite brief, occurring at the end of AGB evolution when photospheric temperature increase during evolution to the proto-planetary nebula phase. Such a wind remnant would therefore be the first to be destroyed by a growing ionised region, resulting in a strong decrease in HI optical depth with increasing ionised radius. Since such a decrease (if it exists at all) does not appear to be profound, we do not favour this explanation as the source of the majority of the observed HI in planetary nebulae. The situation is however less clear if the atomic gas were the product of the photo-dissociation of originally molecular gas by the ultraviolet radiation of the central star, or of the interstellar radiation field. If the interstellar radiation field was responsible for the destruction of the molecular envelope, the atomic component should increase in mass with time, until such time as the molecular gas is depleted. However, many large, evolved planetary nebulae are known to still possess molecular envelopes (e.g. Huggins & Healy 1989), while all HI planetary nebulae are small, making this scenario seem to be unlikely. In the case of photo-dissociation by the central star, one would also expect that the optical depth of HI to grow with time for some period, then decrease as the stellar radiation gets "harder" with increasing photospheric temperatures, and ionisation begins to predominate. Modelling (Gussie et al. 1994) indicates that this is indeed the case, with the optical depth of HI reaching its peak early in PN evolution, when the linear radii of the ionised nebulae is still $\lesssim 0.05$ pc. Since this seems to be consistent with the data presented in figure 4, this explanation is preferred by us. A fourth possibility, that the observed HI is previously ionised material that has since recombined, is considered unlikely since all HI planetary nebulae are young objects, and almost certainly possess growing ionised regions.

Various planetary nebulae display very different brightnesses of CO emission. Only one planetary nebula (NGC 7027) displays $J = 2 \rightarrow 1$ CO emission with an antenna temperature of > 10 K. A few planetary nebulae display $J = 2 \rightarrow 1$ CO emission with antenna temperatures of $\approx$ 1 K, but most planetary nebulae observed have a $J = 2 \rightarrow 1$ CO antenna temperature of $\leq 0.1$ K. It is known that CO emission occurs most commonly in planetary nebulae of Greig's class B or Peimbert's class 1 (Huggins & Healy 1989; Greig 1971; Peimbert 1978). However, the detectability of HI absorption shows no such preference for Greig's B nebulae, as 4 of the 10 nebulae with observed circumnebular HI are of Greig's class C. The Greig classification system (1971) divides forbidden-line dominated, usually bilobate, planetary nebulae (class B), from centrally condensed planetary nebulae (class C). Since Greig's B nebulae are known to be the more massive objects (Kaler 1983), it is suggested that the CO nebulae are often more massive than those nebulae possessing detected circumnebular HI. It should be noted that there are also very few nebulae (IC 5117, BD +30° 3639, and NGC 6302) known to possess *both* a HI and a CO envelope.

One of the nebulae detected in both CO and HI is IC 5117. This nebula is unusual in that it is the one of very few Greig's class C planetary nebula detected in $J = 2 \rightarrow 1$ CO emission. The spatial distribution of the CO and HI envelopes of IC 5117 are however unknown, making a spatial comparison between the molecular and atomic components impossible.

BD +30° 3639 is also detected in $J = 2 \rightarrow 1$ CO emission (Bachiller et al. 1991) and HI emission (TGP). The $J = 2 \rightarrow 1$ CO emission was found to be resolved by the 12″ beam of the IRAM sub-millimetre telescope, although CO cloud was not been mapped by Bachiller et al. The HI cloud of BD +30° 3639 also remains unmapped, but its spatial extent is known to be $\leq 24″$ (Likkel et al. 1992). Such a large maximum size limit does not rule out the possibility that the CO and HI are cospatial. However, the kinematics of the CO and HI gas are very different, making it unlikely the molecular gas and the atomic gas are closely associated. CO emission is detected over a velocity range of 132 km s$^{-1}$, while the velocity range of HI emission is only 54 km s$^{-1}$, identical to the velocity range of the [OIII] emission (Weinberger 1989). The H$_2$ emission of the nebula is also known to extend well beyond the ionised region (Graham et al.

1993). It is therefore considered probably that the HI lies within a larger CO complex, occurring near the edge of the ionised gas in a photodissociation region that shares the kinematic properties of the ionised gas. In this scenario, the CO would be more extended gas that retains the expansion velocity of a fossil stellar wind from a period of relatively high-velocity mass loss, perhaps during the immediate post-AGB phase or slightly later proto-planetary phase.

The third nebula that is observed to possess both a CO envelope (Huggins & Healy, 1989) and a HI envelope (Rodríguez & Moran 1982) is NGC 6302. The distribution of this nebula's HI absorption has been mapped by Rodríguez et al. (1985). The CO envelope is known to extend to > 45" in radius (Nyman, 1989). The extensive distribution of CO is evidenced by the double-peaked nature of the $J = 2 \rightarrow 1$ CO profile and possibly the $J = 3 \rightarrow 2$ CO profile as well. However, no high resolution mapping of the structure able to determine the spatial relationship of CO and HI or HII has been made to the best of our knowledge. However, this nebula is also one of very few planetary nebulae known to possess OH maser emission, which has been mapped by Payne, Phillips, & Terzian (1988). It is found that both the HI absorption and OH maser emission are associated with an optically dark dust lane that bisects the nebula. The OH maser emission is found to be limited to a region of $\leq 10"$, much smaller than the extent of the optical nebula and CO emission. Absorption lines of rotationally excited OH were also observed by Payne, Phillips, & Terzian (1988), who suggest that the ground-state and excited-state OH are cospatial. It is therefore likely that the OH observations only describe a portion of what may be a much larger molecular envelope, the extent of which may be best determined by a more sensitive mapping experiment of the thermally-excited CO emission.

In the three nebulae known to possess both CO and HI, the currently available data is therefore consistent with the hypothesis that the HI is formed within a photodissociation region within a larger molecular envelope and exterior to the ionised gas. However, this may not be universally true, as the nebulae IC 418 is known to possess an extended HI envelope (Taylor, Gussie, & Goss 1990), which is unlikely to be contained within a larger molecular complex. Since IC 418 is not known to possess a CO envelope, in this particular case the photodissociation region may have consumed the entire envelope, possibly with the help of a significant contribution from the ambient interstellar radiation field. IC 418 would then be likely to possess a relatively thin neutral envelope, perhaps indicating that this nebula is of significantly lower mass than IC 5117, BD+30°3639, and NGC 6302.

It is therefore suggested by the preponderance of massive B class nebulae among those nebulae known to possess CO that CO is rapidly destroyed (or possibly never formed) in the envelopes of low-mass planetary nebulae, but often (albeit not universally) manages to form and survive in the envelopes of more massive nebulae. This is supported by a theoretical model by Mamon, Glassgold, and Huggins (1988) who model the photodissociation of CO in a circumstellar envelope of a late-type star due to UV radiation from the interstellar radiation field. They find that CO in low mass-loss rate envelopes ($\dot{M} < 10^{-5} M_\odot$ yr$^{-1}$) is rapidly destroyed, resulting in an optically thin CO cloud with linear radii less than $R_{CO} = 0.1$ pc, while envelopes with higher mass-loss rates ($\dot{M} > 10^{-5} M_\odot$ yr$^{-1}$) are optically thick and extend to $R_{CO} \approx 1$ pc. Since the ionised linear radii of the planetary nebulae are typically $R_{ion} \approx 0.1$ pc, it can easily be seen that the truncated, optically thin CO clouds of low mass-loss rate planetary nebulae would be rapidly destroyed at the onset of ionisation, while higher mass-loss rate nebulae would still possess an appreciable CO envelope.

If CO is actively destroyed in planetary nebulae, neutral carbon atoms (CI) should form. Observations by Beichman et al. (1983) have detected CI in the proto-planetary nebula CRL 2688 using the $v = 492$ GHz fine-structure line. However, they failed to detect CI in seven other stars. CI has only recently been discovered in bonafide planetary nebula by Bachiller et al. (1994). The apparent difficulty in detecting CI can be explained by the fact that the ground-state dissociation energy of the CO molecule is $D_0 = 11.09$ eV, only slightly lower than the ionisation potential of the carbon atom ($I = 11.26$ eV), thus making the subsequent ionisation of the dissociated carbon atoms relatively rapid. The photochemical model of Mamon, Glassgold, & Huggins (1988) confirm that the CI are quickly ionised to form CII. One would then expect to that ionised carbon would be cospatial with HI or $H_2$ material in planetary nebulae. This has yet to be ascertained, as cold CII is only visible as a fine-structure line at $\lambda = 158 \mu$m; a wavelength that is inaccessible from the ground. Airborne telescopes have observed this line in PNe (e.g. Dinerstein, Haas, & Werner, 1991), but their spatial resolution is insufficient to map the planetary nebula envelopes. The next generation of airborne observatories should make an effort to observe CII in planetary nebulae.

7. SUMMARY AND CONCLUSION

Circumnebular HI is detected in the planetary nebulae SwSt 1 and M 3-35, and is tentatively detected in Hu 2-1 as well. HI absorption is also detected in M 1-78, and the velocity of the HI absorption line of M 1-78 is consistent with CO emission in the direction to the nebula. It remains unclear whether or not the observed neutral gas is causally connected with the nebula, but since M 1-78 is now believed to be a HII region, the neutral gas may well lie within the progenitor cloud.

$J = 2 \rightarrow 1$ CO is detected in the planetary nebula M 1-59, and previous detections of $J = 2 \rightarrow 1$ CO emission in IC 5117 and NGC 6302 are confirmed. $J = 3 \rightarrow 2$ CO emission is detected in IC 5117 and NGC 6302.

The new data are consistent with the existence of an inverse correlation between the optical depth of circumnebular HI and the linear radius of the ionised nebula, but the data are still to few to confirm that the correlation is statistically significant. It therefore remains unknown whether the ionised shell in planetary nebulae grows at the expense an outer atomic shell. There consequently remains three viable explanations for the formation of circumnebular HI; the photo-dissociation of molecular gas by ultraviolet radiation from the central star, photo-dissociation of molecular gas by interstellar ultraviolet radiation, and an atomic mass-loss phase during the late AGB evolution of the stellar progenitor.

8 ACKNOWLEDGMENTS

This research was supported by a grant from the Natural Sciences and Engineering Research Council of Canada and a grant from the Australian Research Council. This research has made use of the SIMBAD database, operated by CDS, Strasbourg, France.

TABLE 1

Results of the HI Observations

| Name | PK# | $\lambda = 21$cm Flux mJy | Peak $\tau$ | $\int \tau dv$ km s$^{-1}$ | $v_{exp(HI)}$ km s$^{-1}$ |
|---|---|---|---|---|---|
| SwSt1 | 001–06°2 | $27.3 \pm 3.3$ | $0.67 \pm 0.13$ | $8.0 \pm 0.6$ | $18.6 \pm 2.3$ |
| NGC 6567 | 011–00°2 | $150 \pm 20$ | $\leq 0.069$ | | |
| M 2-43 | 027+04°1 | $22.0 \pm 4.7$ | $\leq 0.24$ | | |
| NGC 6741 | 033–02°1 | $32.0 \pm 4.0$ | $\leq 0.065$ | | |
| Hu 2-1 | 051+09°1 | $46 \pm 2$ | $0.18 \pm 0.04$ | $0.68 \pm 0.09$ | $5 \pm 2$ |
| M 3-35 | 071–02°1 | $31.8 \pm 1.7$ | $0.16 \pm 0.06$ | $2.5 \pm 0.3$ | $22.0 \pm 3.5$ |
| NGC 6884 | 082+07°1 | $145 \pm 3$ | $\leq 0.048$ | | |
| IC 2165 | 221–02°1 | $160 \pm 10$ | $\leq 0.023$ | | |
| M 1-26 | 358–00°2 | $118 \pm 4$ | $\leq 0.090$ | | |

Notes:

1) Error bars in $\tau$ represent the $1\sigma$ noise of the spectra.

2) Upper limits in $\tau$ represent the $3\sigma$ noise of the spectra.

TABLE 2

Results of the $J = 2 \rightarrow 1$ CO Observations

| Name | PK# | Peak $T_{MB}$ K | $\int T_{MB}\,dv$ K km s$^{-1}$ | $v_{LSR(CO)}$ km s$^{-1}$ | $v_{exp(CO)}$ km s$^{-1}$ |
|---|---|---|---|---|---|
| SwSt 1 | 001–06°1 | $\leq 0.047$ | | | |
| NGC 6644 | 008–07°2 | $\leq 0.042$ | | | |
| M 1-59 | 023–02°1 | $0.038 \pm 0.009$ | $0.50 \pm 0.08$ | $114 \pm 1$ | $13 \pm 2$ |
| M 2-43 | 027+04°1 | $\leq 0.030$ | | | |
| NGC 6741 | 033–02°1 | $\leq 0.078$ | | | |
| NGC 6790 | 037–06°1 | $\leq 0.048$ | | | |
| Hu 2-1 | 051–09°1 | $\leq 0.048$ | | | |
| IC 4997 | 058–10°1 | $\leq 0.40$ | | | |
| NGC 6886 | 060–07°2 | $\leq 0.057$ | | | |
| BD+30°3639 | 064+05°1 | $\leq 0.20$ | | | |
| M 3-35 | 071–02°1 | $\leq 0.036$ | | | |
| NGC 6884 | 082+07°1 | $\leq 0.048$ | | | |
| IC 5117 | 089–05°1 | $0.5 \pm 0.1$ | $6 \pm 1$ | $-8 \pm 1$ | $17 \pm 2$ |
| IC 418 | 215–24°1 | $\leq 0.19$ | | | |
| IC 2165 | 221–12°1 | $\leq 0.060$ | | | |
| M 1-16 | 226+05°1 | $0.97 \pm 0.08$ | $23.0 \pm 0.8$ | $42 \pm 1$ | $27 \pm 3$ |
| M 1-11 | 232–04°1 | $\leq 0.15$ | | | |
| NGC 3918 | 294+05°1 | $\leq 0.051$ | | | |
| IC 972 | 326+42°1 | $\leq 0.11$ | | | |
| NGC 6302 | 349+01°1 | $0.67 \pm 0.04$ | $13.5 \pm 0.3$ | $-35 \pm 1$ | $28 \pm 4$ |
| H 1-35 | 355–03°3 | $\leq 0.13$ | | | |

Notes:

1) Error bars in $T_{MB}$ represent the $1\sigma$ noise of the spectra.

2) Upper limits in $T_{MB}$ represent the $3\sigma$ noise of the spectra.

TABLE 3

Results of the $J = 3 \rightarrow 2$ CO Observations

| Name | PK# | Peak $T_{MB}$ K | $\int T_{MB}\, dv$ K km s⁻¹ | $v_{LSR(CO)}$ km s⁻¹ | $v_{exp(CO)}$ km s⁻¹ |
|---|---|---|---|---|---|
| SwSt1 | 001−06°1 | ≤ 1.2 | | | |
| NGC 6644 | 008−07°2 | ≤ 0.87 | | | |
| M 1-59 | 023−02°1 | ≤ 0.84 | | | |
| M 2-43 | 027+04°1 | ≤ 0.69 | | | |
| NGC 6741 | 033−02°1 | ≤ 0.90 | | | |
| NGC 6790 | 037−06°1 | ≤ 0.69 | | | |
| NGC 6803 | 046−04°1 | ≤ 0.63 | | | |
| Hu 2-1 | 051+09°1 | ≤ 1.0 | | | |
| IC 4997 | 058−10°1 | ≤ 0.69 | | | |
| NGC 6886 | 060−07°2 | ≤ 0.41 | | | |
| M 3-35 | 071−02°1 | ≤ 0.81 | | | |
| Hu 1-2 | 086−08°1 | ≤ 0.63 | | | |
| IC 5117 | 089−05°1 | 1.5 ± 0.3 | 8.7 ± 0.2 | -16 ± 2 | 12 ± 2 |
| NGC 6302 | 349+01°1 | 3.8 ± 0.9 | ≥ 8 | ? | ? |
| H 1-35 | 355−03°3 | ≤ 0.84 | | | |

Notes:

1) Error bars in $T_{MB}$ represent the $1\sigma$ noise of the spectra.

2) Upper limits in $T_{MB}$ represent the $3\sigma$ noise of the spectra.

3) The spectrum of NGC 6302 is too noisy to determine the velocity limits of the emission.

TABLE 4

Compilation of Planetary Nebulae Known to Possess HI

| Name | PK# | Distance kpc | Radius " | Radius pc | $\tau_{peak}$ | $\int \tau\, dv$ km s$^{-1}$ |
|------|-----|--------------|----------|-----------|---------------|------------------------------|
| SwSt 1 | 001–06°1 | 1.2 [12] | 0.65 [10] | 0.0038 | 0.67±0.13 | 8.0±0.6 |
| NGC 6790 | 037–06°1 | 1.6 [4] | 1.8 [10] | 0.014 | 0.12±0.02[1] | 1.50±0.12[1] |
| Hu 2-1 | 051+09°1 | 1.5 [12] | 0.9 [10] | 0.0065 | 0.018±0.004 | 0.68±0.09 |
| IC 4997 | 058–10°1 | 2.6 [2] | 0.8 [8] | 0.010 | 0.13±0.02[2] | 2.1±0.3[3] |
| NGC 6886 | 060–07°2 | 3.0 [4] | 3.0 [10] | 0.044 | 0.05±0.01[1] | 0.87±0.12[1] |
| BD+30°3639 | 064+05°1 | 2.8 [5] | 2.5 [8] | 0.011 | 0.03±0.01[1] | 0.2-2.0[1] |
| M3-35 | 071–02°1 | 2.4 [11] | 2.0 [11] | 0.023 | 0.16±0.06 | 2.5 ± 0.3 |
| IC 5117 | 089–05°1 | 2.1 [4] | 1.0 [10] | 0.011 | 0.01±0.03[1] | 2.10±0.14[1] |
| IC 418 | 215–24°1 | 1.0 [6] | 6.2 [9] | 0.030 | 0.01±.002[1] | 0.16±0.03[1] |
| NGC 6302 | 349+01°1 | 0.7 [11] | 4.5 [7] | 0.015 | 0.22±0.02[3] | 3.5±0.4[3] |

References:

1) Taylor, Gussie, & Pottasch (1990)

2) Altschuler et al. (1986)

3) Rodríguez & Moran (1982)

4) Sabbadin (1984)

5) Masson (1989)

6) Taylor & Pottasch (1987)

7) Rodríguez et al. (1985)

8) Pottasch (1984)

9) Aaquist & Kwok (1990)

10) Maciel (1984)

11) Weinberger (1989)

FIGURE CAPTIONS

FIGURE 1. The recorded HI absorption spectra of the survey nebulae detected in HI absorption. The HI detected in M 1-78 is thought to be interstellar in origin. See text.

FIGURE 2: $J = 2 \rightarrow 1$ CO spectra of surveyed planetary nebulae with detected emission.

FIGURE 3: $J = 3 \rightarrow 2$ CO spectra of surveyed planetary nebulae with detected emission.

FIGURE 4. The integrated optical depth of circumnebular atomic hydrogen as a function of linear radius for all nebulae detected in HI absorption. Error bars are not included for the $x$-axis since they are difficult to estimate. The diagonal dashed lines are theoretical lines for the model atomic winds of TGP with $R_{HI} \gg R_{ion}$, expansion velocities of $v_{exp} = 10 \, \mathrm{km \, s^{-1}}$, excitation temperatures of $T_{HI} = 100 \, \mathrm{K}$, mean molecular weights of $\mu = 1.4$, and with differing atomic mass-loss rates of $\dot{M} = 2 \times 10^{-5} M_{\odot} \mathrm{yr^{-1}}$ and $\dot{M} = 10^{-6} M_{\odot} \mathrm{yr^{-1}}$ for the upper and lower lines, respectively.

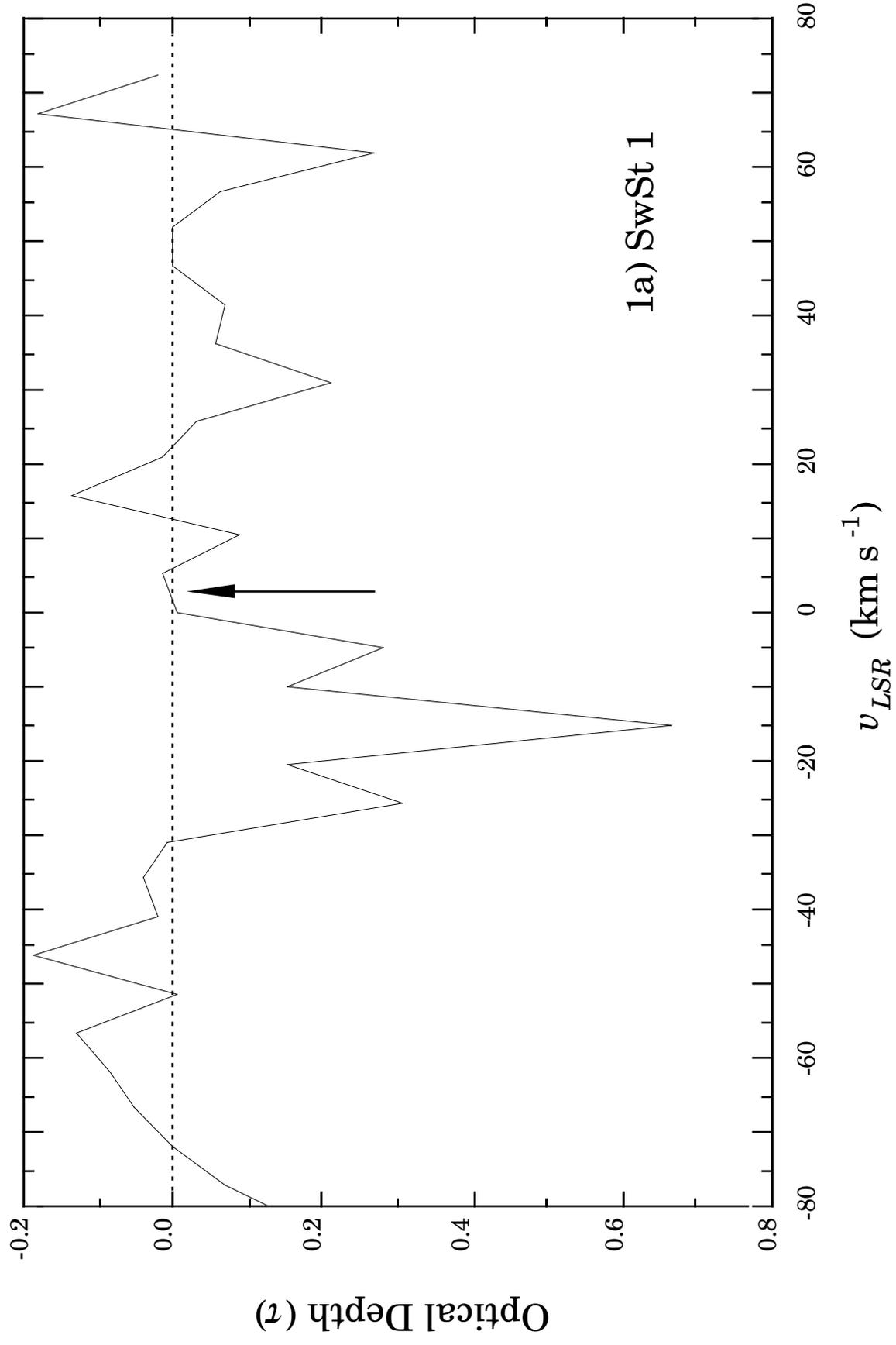

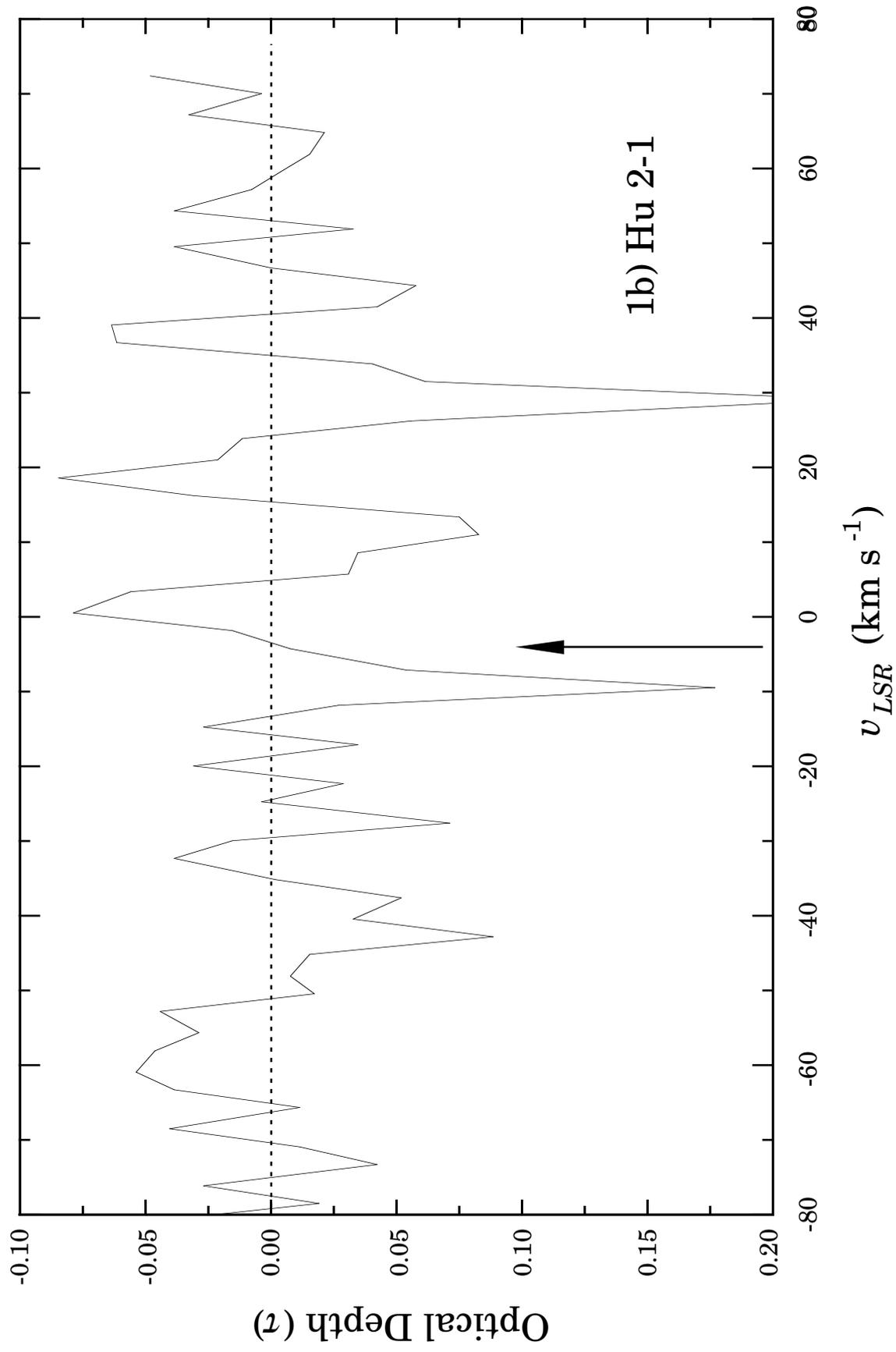

1b) Hu 2-1

$v_{LSR}$ (km s$^{-1}$)

Optical Depth ($\tau$)

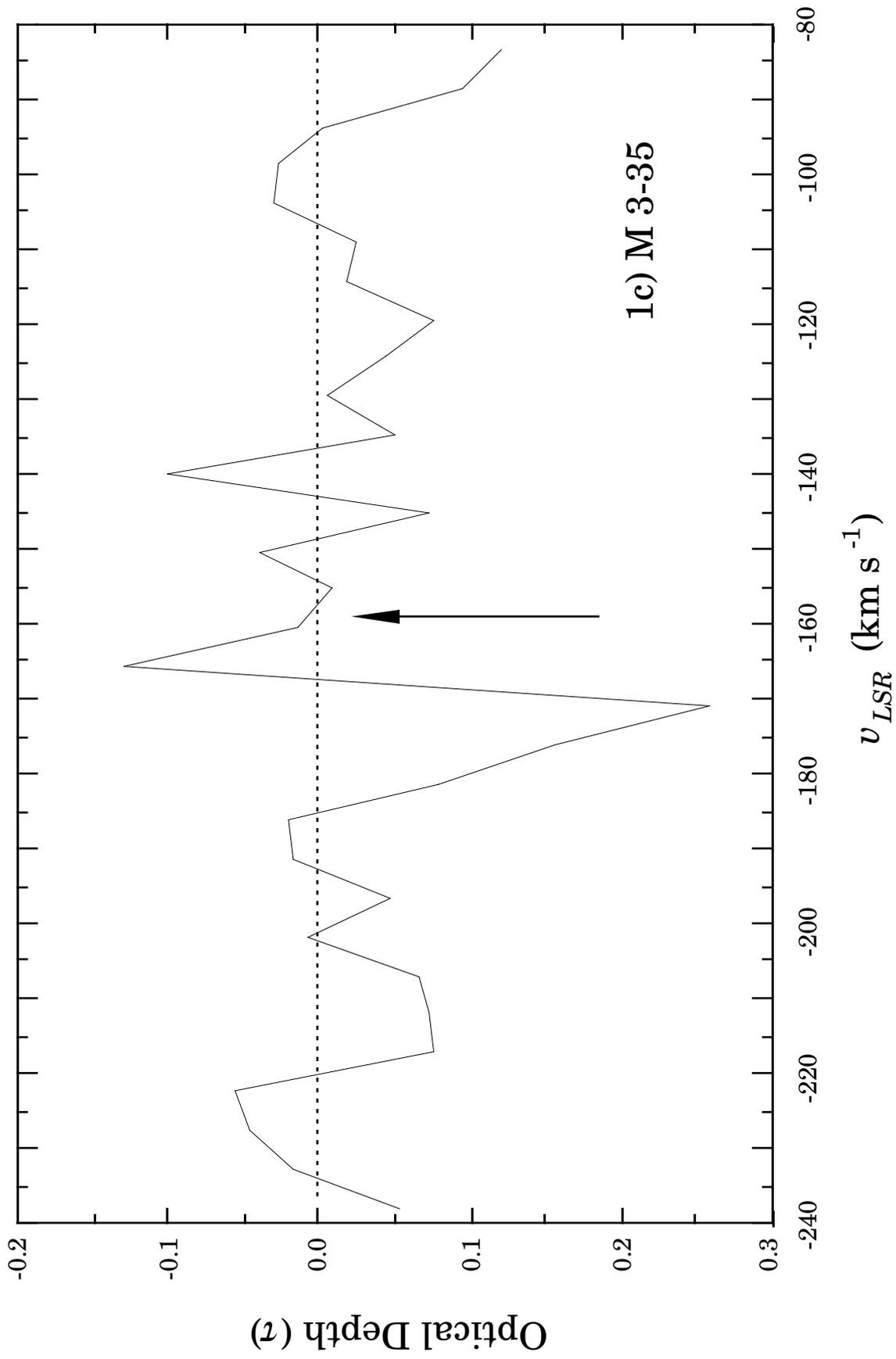

1c) M 3-35

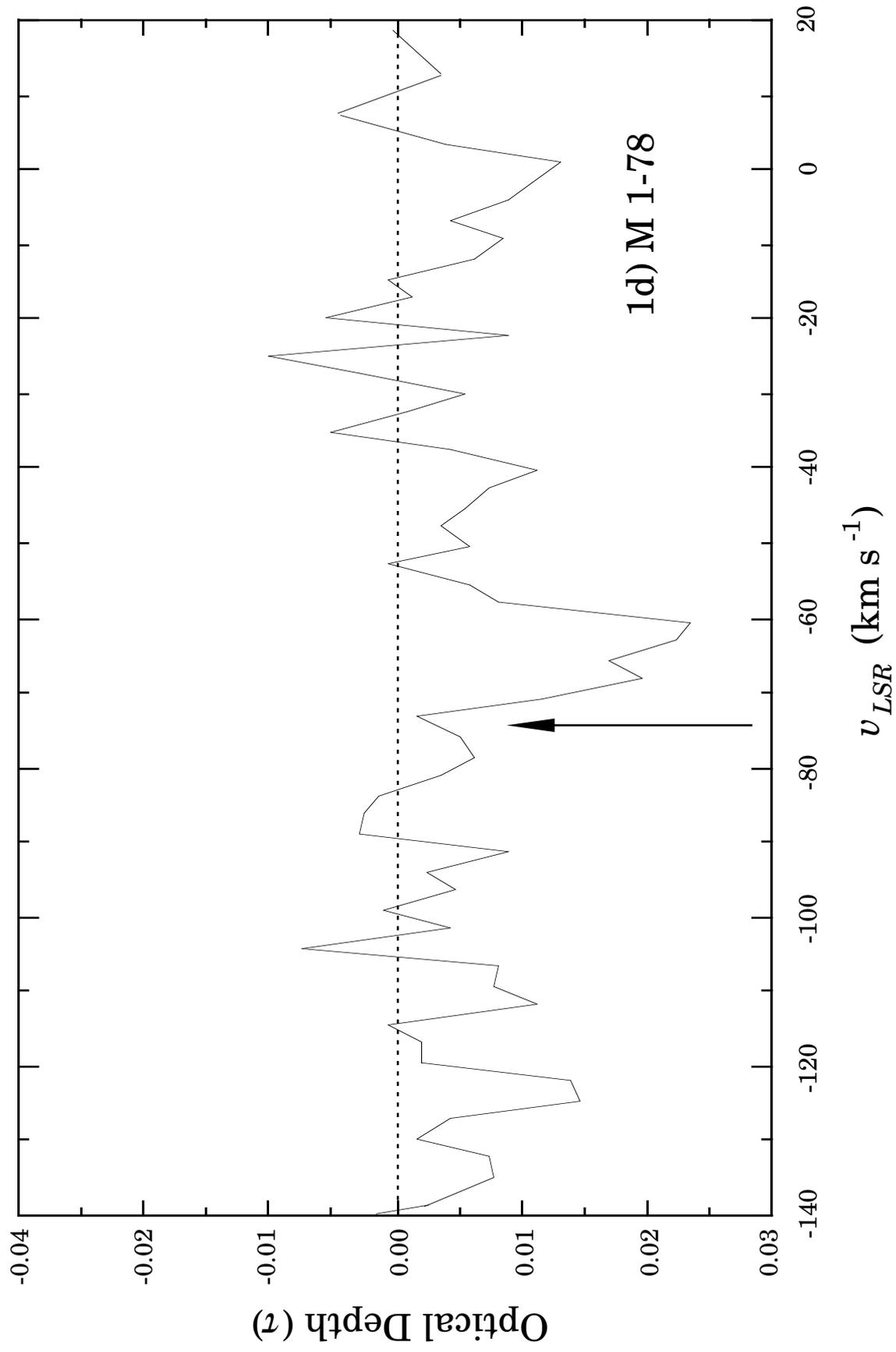

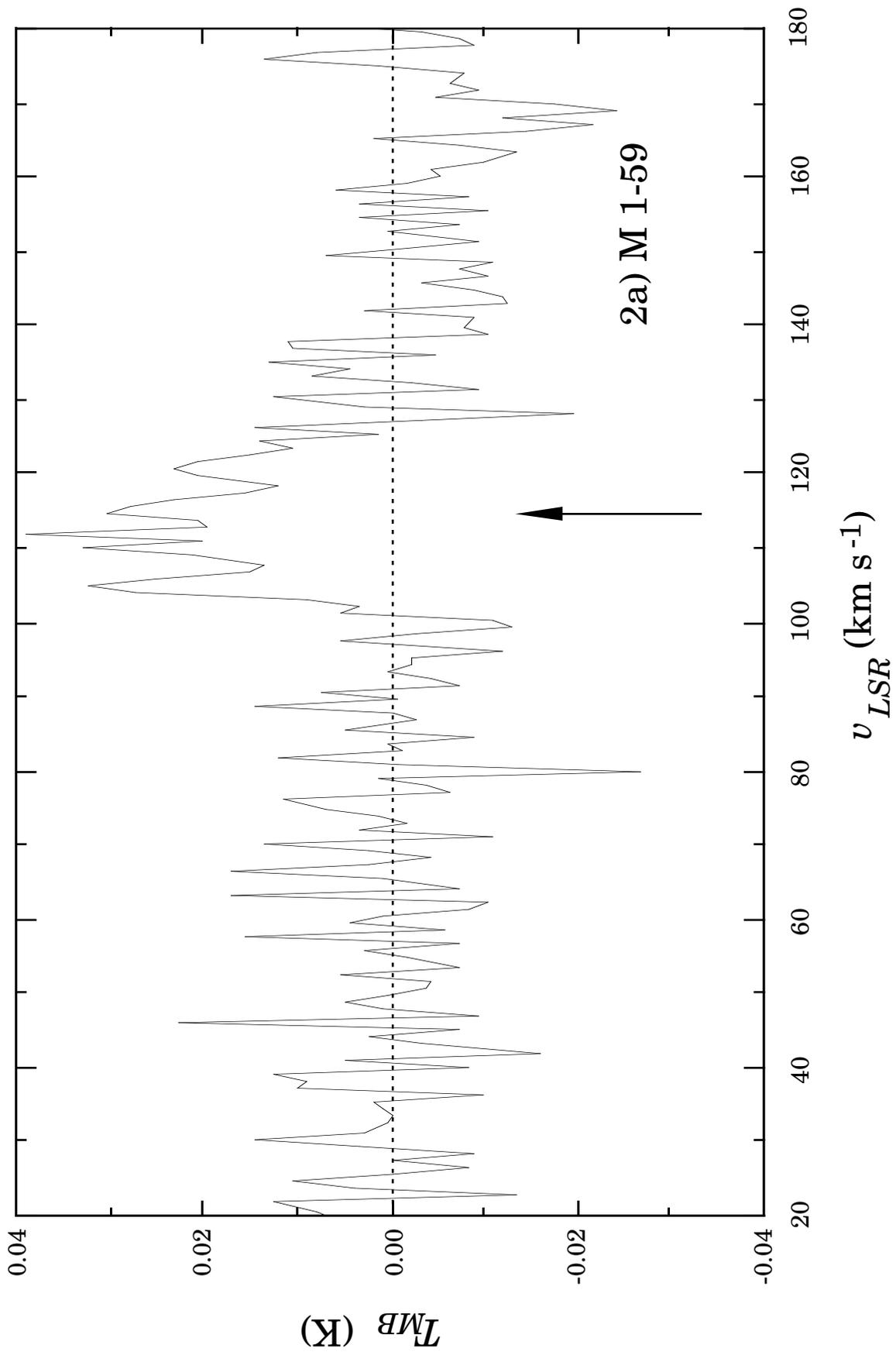

2a) M 1-59

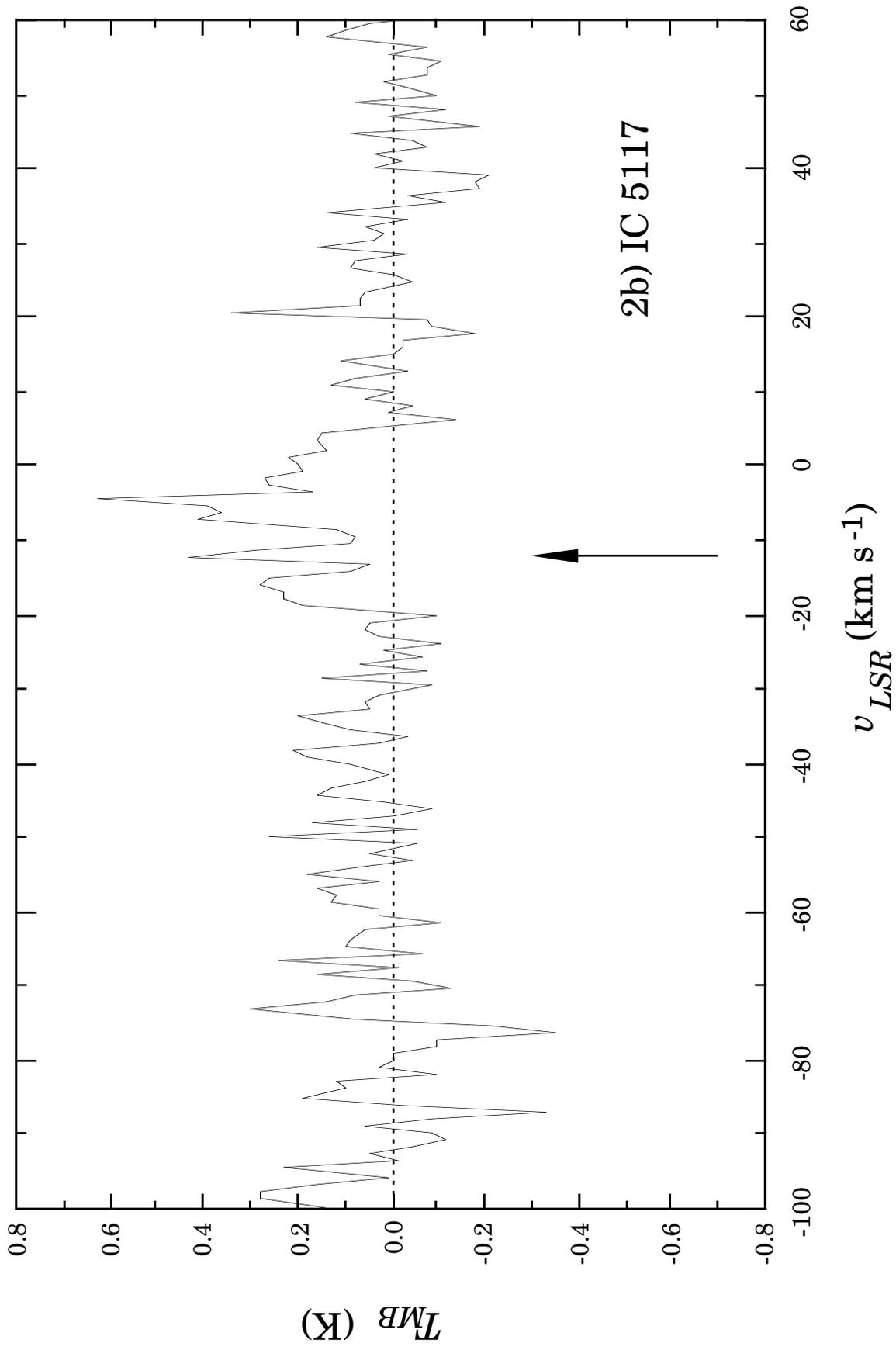

2b) IC 5117

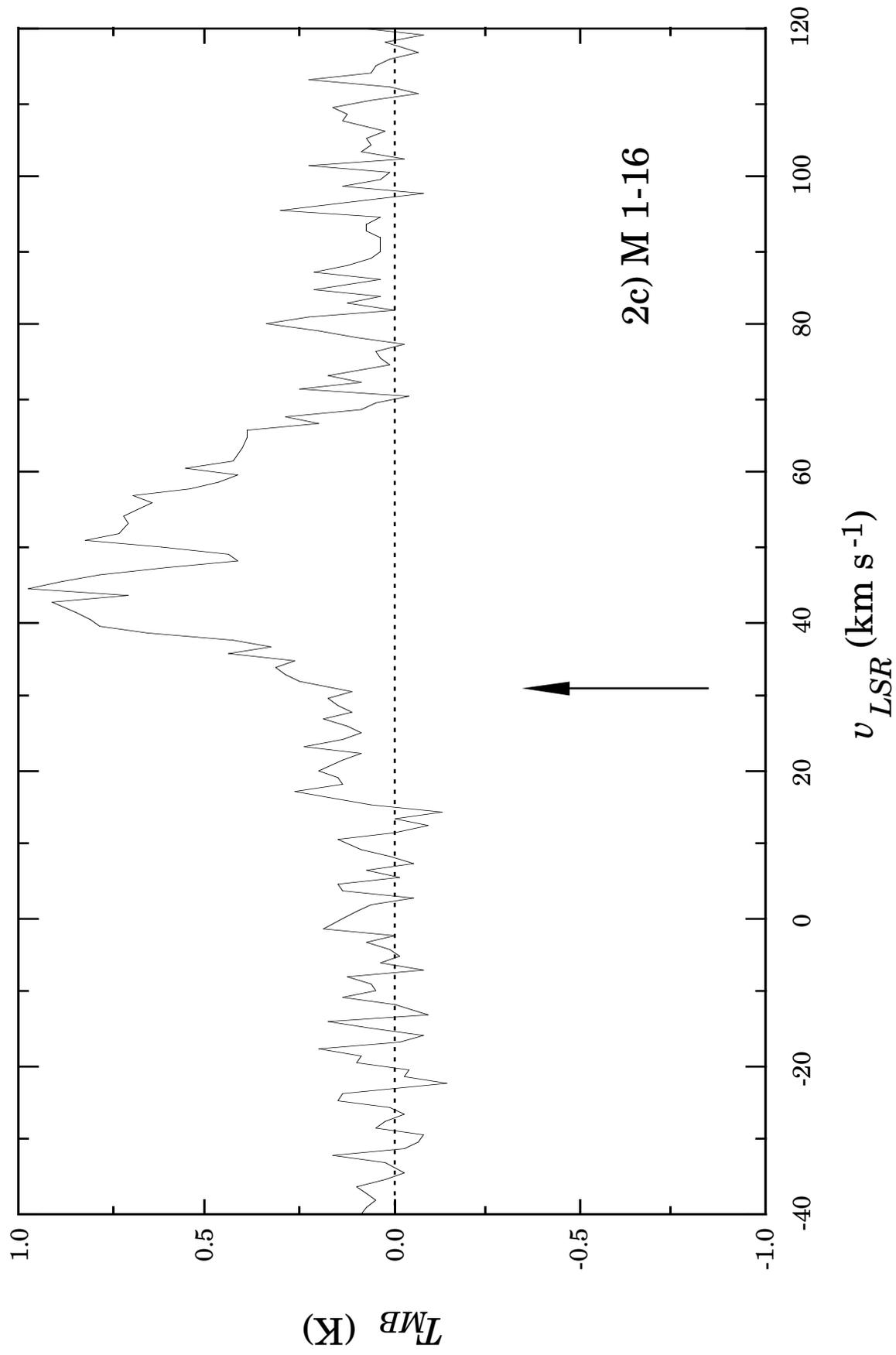

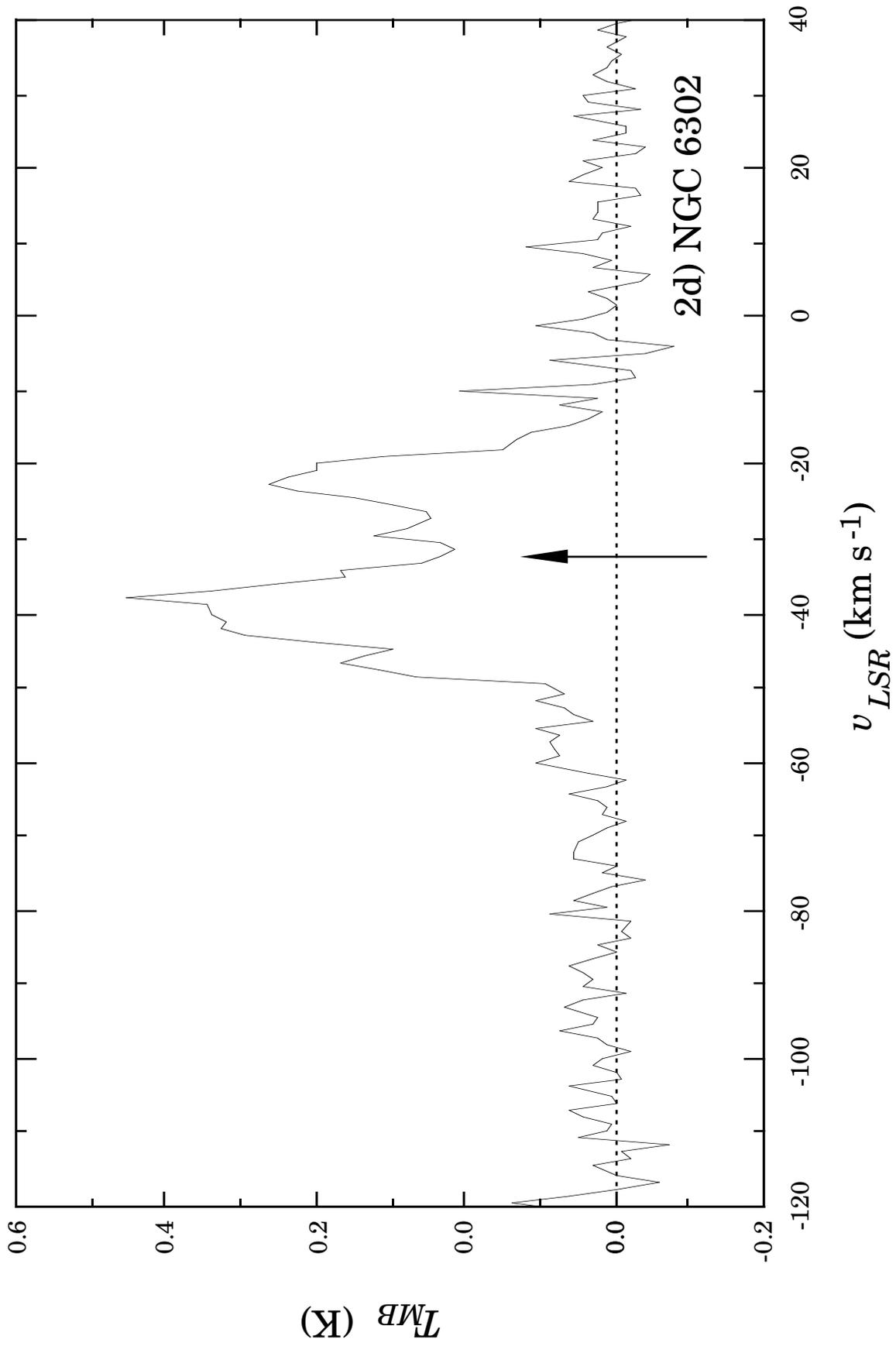

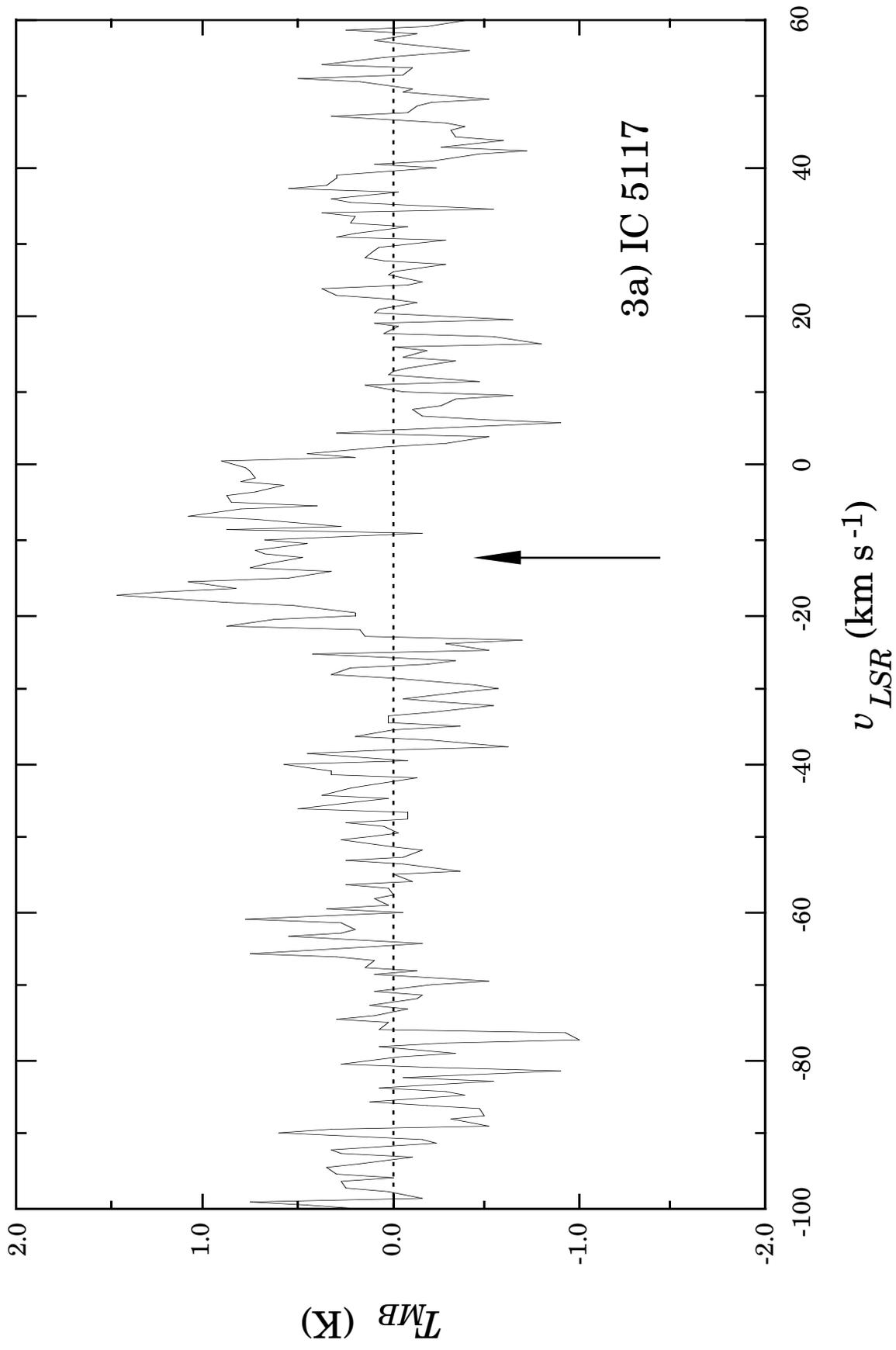

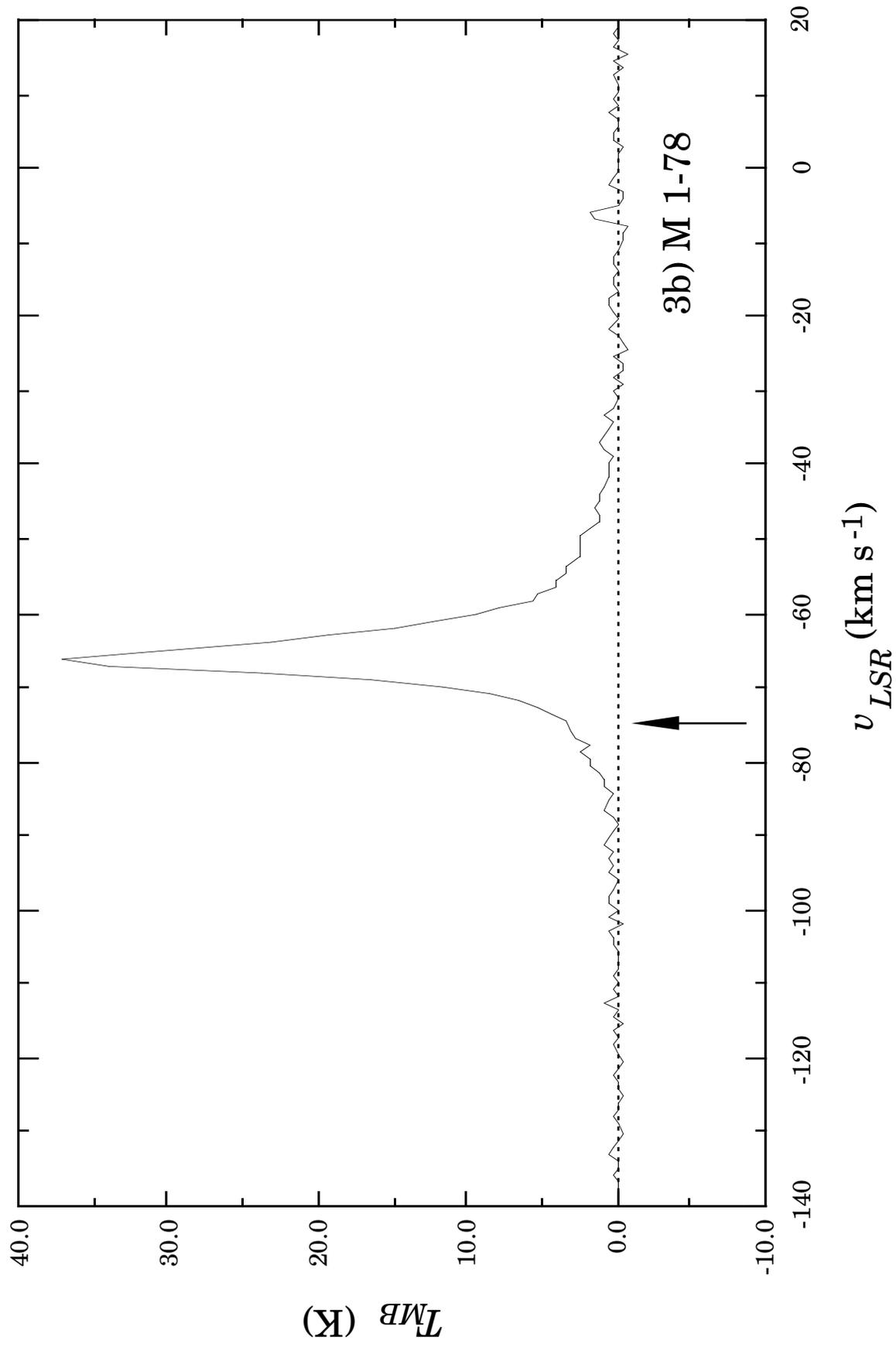

3b) M 1-78

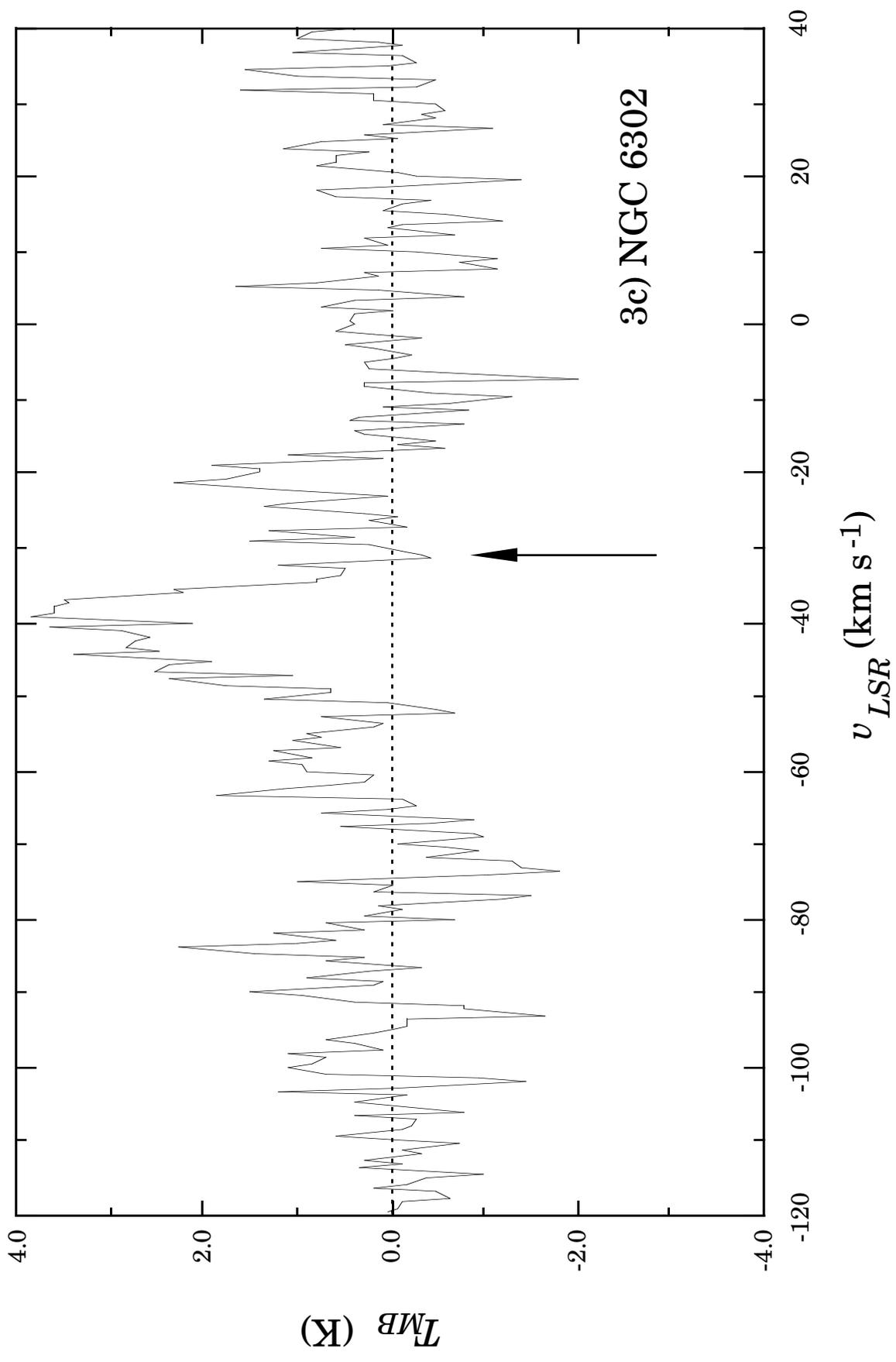

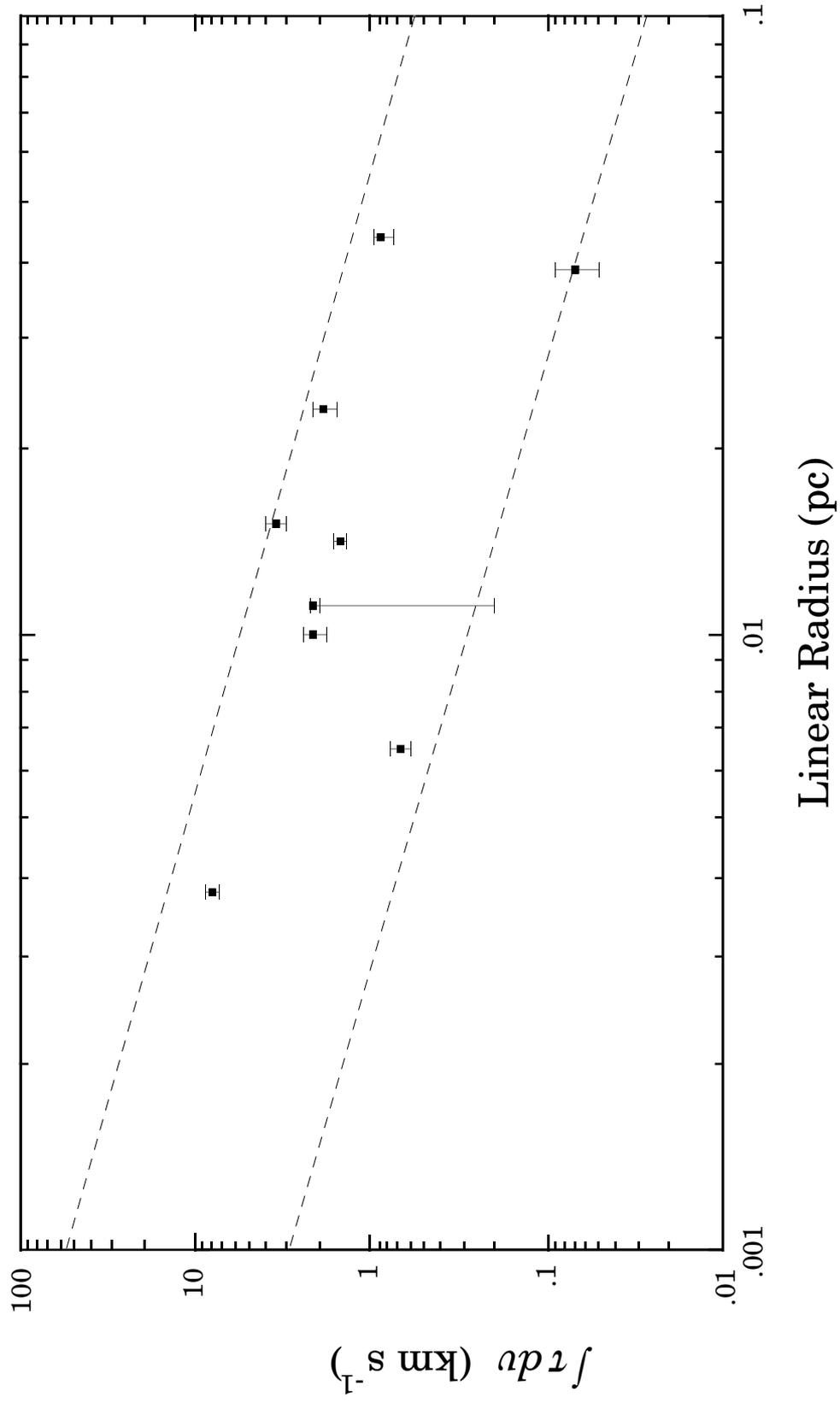

Figure 4